# RFQ PARAMETER CHOICE BY MULTI-PARAMETER OPTIMIZATION

Boris I. Bondarev, Alexander P. Durkin
*Moscow Radiotechnical Institute RAS,*
132, Warshavskoe shosse, Moscow, Russia, 113519
lidos@aha.ru

Experts in different accelerator laboratories accumulated wide experience in RFQ designing. Nevertheless new tasks on high-power, high-energy CW linac designing (for example, for ADT applications) were appeared recently. Such accelerators must be practically free of beam losses. This main condition placed more stringent requirements upon beam quality at the input of each linac part. Because RFQ is used as beginning part in above accelerators beam high-quality requirements is related to RFQ primarily. It means that existing experience in RFQ designing must be enriching by new approaches and new methodology.

Both acceleration and focusing inside RFQ are governed by smooth variation of four parameters: amplitude of vane modulation, intervane voltage, synchronous phase and bore aperture. Experience shows that RFQ channel can be arbitrarily divided into three parts: adiabatic buncher (where beam is bunching and accelerated by small accelerating fields and synchronous phase near to $-90^0$), shaper (where smooth transition to nominal accelerating field and nominal synchronous phase take place) and proper accelerator. Lengths of RFQ parts and RFQ parameter variations are determined by the trial-and-error method. The process is labor-intensive and depends on a choice of the first approximation.

The mathematical methods of many-parametric optimization (as applied to charge particle beam dynamics) were developed at present time. In principle, they give a possibility to solve problem of RFQ optimization.

The principle of mathematical optimization consists in choice of optimization criteria, choice of control parameters and choice of exhaustive method. The possible criteria of optimization for RFQ channel designing can be: minimal length of RFQ channels and beam transmission no less then given one. Vane modulation, intervane voltage, synchronous phase and bore aperture in each cell can be used as control parameters. But such optimization method is possible in principle but it cannot be realized in practice. If we make an attempt to realize optimization procedure in form described above a failure is guaranteed. Time needed to solve the procedure will be astronomical large because control parameter number are more then thousand. It must be added that time for one RFQ version simulation is measured by minutes. Modern computer techniques have no power to make such optimization.

It means that simplified model for RFQ beam acceleration and focusing is needed. This model must, from one side, hardly reduce control parameter number and, from another side, give equivalent description of beam motion. Below we propose a model that allows to separate longitudinal motion optimization from transverse motion one.

Let us consider beam longitudinal motion. It declares by following equation

$$\frac{d^2z}{d\tau^2} = \frac{eUk\sigma}{2W_0} I_0(kr)\cos(kz)\cos(\omega\tau + \varphi_s), \quad (1)$$

where $\tau = ct$, $U$ - is intervane voltage, $k = 2\pi/L$, $L$ is modulation period, $e$ and $W_0$ are particle charge and rest energy, $\omega = 2\pi/\lambda$, $\lambda$ is RF wave length, $r$ is distance between particle and channel axis, $\sigma = \frac{m^2 - 1}{m^2 I_0(ka) + I_0(mka)}$. The non-relativistic approximation $\gamma \cong 1$ is used.

Let us resolve RF field into two travelling harmonics and average over harmonics $\omega\tau + kz$, we receive an equation

$$\frac{d^2z}{d\tau^2} = \frac{eUk\sigma}{4W_0}\cos(\omega\tau - kz + \varphi_s).$$

For synchronous particle $\omega\tau + kz$. Let us introduce variable $y = \omega\tau - kz = k(z - z_s)$. Taking $L = \beta_s\lambda$ and using equations for $\beta_s$ and $z - z_s$, we obtain

$$y''_\tau + \frac{2L'_\tau}{L}y'_\tau + \frac{L''_\tau}{L}y - \Omega^2\left(\cos\varphi_s - \cos(\varphi_s + y)\right) - \frac{ek^2}{W_0}\frac{\partial U_c}{\partial y} = 0,$$

where $\Omega^2 = \frac{4\pi eUT}{W_0L^2}$, $T = \frac{\pi}{4}\sigma$, $U_c$ is Coulomb field potential. Using relations

$$\frac{L'}{L} = \frac{\Omega^2}{\omega}\cos\varphi_s,$$

$$\frac{L''_\tau}{L} = \frac{\Omega^2}{\omega}\frac{(UT)'_\tau}{(UT)}\cos\varphi_s + \frac{\Omega^2}{\omega}(\cos\varphi_s)'_\tau - \left(\frac{\Omega^2}{\omega}\cos\varphi_s\right)^2$$

and taking $\Omega^2 = \Omega_0^2 \eta \frac{W_n}{W}$, where $W_n$ and $W$ are input and current energies of particles, $\Omega_0^2 = \frac{2\pi e(UT)_{max}}{W_n\lambda^2}$, $\eta = \frac{UT}{(UT)_{max}}$. If we convert to independent variable $\zeta = \Omega_0\tau$, we finally obtain

$$y'' + 2\frac{\Omega_0}{\omega}h\frac{W_n}{W}\cos\varphi_s \, y' + \left(\frac{\Omega_0}{\omega}\frac{W_n}{W}\cos\varphi_s \, h' + \frac{\Omega_0}{\omega}h\frac{W_n}{W}(\cos\varphi_s)' - (\frac{\Omega_0}{\omega}h\frac{W_n}{W}\cos\varphi_s)^2\right)y -$$
$$- h\frac{W_n}{W}(\cos\varphi_s - \cos(y+\varphi_s)) - \frac{ek^2}{W_0\Omega_0^2}\frac{\partial U_c}{\partial y} = 0 \qquad (2)$$

Equation (2) together with synchronous particle equation

$$(\frac{W}{W_n})' = 2\frac{\Omega_0}{\omega}h\cos\varphi_s \qquad (3)$$

determines longitudinal motion in RFQ channel in equivalent travelling wave approximation.

If optimization is made only by Eqs (2) and (3), then final results can be obtained with high defocusing parameters and very small transverse acceptance. It means that restriction on defocusing parameter must be added to Eqs (2) è (3)

$$A_{def} = 2(\frac{\Omega_0}{\omega})^2 h\frac{W_n}{W}|\sin\varphi_s| \leq A \qquad (4)$$

Parameter A usually lies in the range of A = 0.01 – 0.015.

In travelling wave approximation we see that longitudinal motion is determined only by one given parameter $k = \Omega_0/\omega$ and by laws of $h$ and $\varphi_s$ variations.

We propose to calculate longitudinal component of Coulomb field as ones for cylinder with constant radius and with uniform distribution in each transverse crossection. In longitudinal direction periodical modulation take place along axis Z. In this case each macroparticle can be presented as uniform charge disc. One hundred discs are enough for reliable estimates.

If $\tilde{U}$ is Coulomb field potential with charge unit in beam, then Coulomb term in Eq. (3) has a form $\alpha(\frac{W_n}{W})^{3/2}\frac{\partial \tilde{U}}{\partial y}$, where $\alpha = \frac{I}{\varepsilon_0 c \beta_n (UT)_{\max}}$.

If high-current beam is accelerated, then this beam must not be pinched in longitudinal direction in optimal mode. In this case new limitation that demands monotonic variation of rms beam width $\langle\Delta\varphi\rangle^2$. In ideal this limitation looks as

$$\frac{d\langle\Delta\varphi\rangle^2}{dz} \leq 0 \qquad (5)$$

In order to finish definition of optimization task for longitudinal motion it is needed to parameterize functions $h(z)$ and $\varphi_s(z)$. As far as there are no oscillations with frequency $\omega$ in external force it is enough to define several plot points and intermediate point values determine by interpolation (linear in the simplest cases). In practice, 20 points is enough.

So longitudinal motion optimization can be reduced to search values of functions $h(z)$ and $\varphi_s(z)$ in number of extraction points in order to solve Eqs (2) and (3) with limitations (4) and (5) and obtain version satisfied given criteria.

Let us consider transverse motion optimization. Let us suppose that longitudinal motion optimization by above scheme was successful. As a result we obtain functions $h(z)$ and $\varphi_s(z)$. Using relations between variable $z$ and cell number $n = 2z/k$, we obtain for each cell values $U(n)T(n)$, $\varphi_s(n)$, $A_{def}(n)$ and factor of density increasing by beam bunching $q(n)$ connected with rms values of beam width. Now we are needed to find optimal values of intervane voltage $U(n)$ and aperture $a(n)$ or mean radius $R_0(n)$.

It is known that focusing period in RFQ is defined by three parameters: focusing parameter $\hat{A}$ (proportional to $U/R_0^2$), defocusing parameter $A_{def}$ and Coulomb parameter proportional to $qI/\varepsilon$. Here $R_0$ is value near to mean radius, $I$ is beam current, $\varepsilon$ is beam emittance. For ideal form of vane shape - $R_0^2 = a^2/(1-\sigma I_0(ka))$. Each parameter triplet (if it placed inside stability zone) is corresponded to matched radius $r$ maximal over period. Relation $E = \frac{R_0^2}{r^2}\varepsilon$ defines transverse channel acceptance. We know values for defocusing and Coulomb factors obtained by longitudinal motion calculation. Because RF field intensity is limited by some value then value $U/R_0$ is also limited by $E_b$. Using maximal value we obtain that focusing factor is proportional to $E_b/R_0$. Both values $r$ and $R_0$ are decreased with focusing parameter growth. Value $\tilde{R}_0(n)$ corresponds to maximal acceptance and is quest optimal value.

By next step we find $U(n) = E_b\tilde{R}_0(n)$, $\tilde{T}(n) = (UT)(n)/U(n)$, and find $m(n)$ and $a(n)$ on base of $\hat{O}$ and $R_0$.

Transverse motion calculations were not needed if we use proposed optimization method. Instead, it is enough to find periodical solution for channel envelope in each cell.

So proposed method gives a possibility to use mathematical methods of multi-parametric optimization. Only values of two parameters in chosen points must be found and only longitudinal motion equations (2), (3) must be solved with limitations (4), (5), using only small number of particles and fast calculation of Coulomb attraction. Optimization was based on simple model but main components of external forces and linear component of Coulomb forces are taken into account. Of course it cannot be advocated that optimal solution will be kept when other factors will be taken into account. Nevertheless it can be assumed that it will be near to optimal. If obtained optimal solution will be used as initial approximation then procedure of further "hand" optimization will be not time consumed.

At present time code package for RFQ optimization is generated in the frame of ISTC Project # 912. In future it will insert into LIDOS.RFQ.Designer as its part [1].

If using of proposed optimization method will be successful then it will be possible to generate data base for optimal functions $h(z)$ and $j_s(z)$ and each twosome of $k$ and $a$ values. In this case initial approximation choice will be simpler.

## REFERENCES

[1] B.I.Bondarev, A.P.Durkin, S.V.Vinogradov, I.V.Shumakov "New Tasks and New Codes for RFQ Beam Simulation", this Conference.